# Preparation, structural, dielectric and magnetic properties of $LaFeO_3$-$PbTiO_3$ solid solutions

S. A. Ivanov[a,b], R. Tellgren[c], F. Porcher[d], T. Ericsson[e], A. Mosunov[a], P. Beran[f], S. K. Korchagina[a], P. Anil Kumar[b], R. Mathieu[b,*], P. Nordblad[b]

a -Department of Inorganic Materials, Karpov' Institute of Physical Chemistry, Moscow, Russia
b -Department of Engineering Sciences, Box 534, Uppsala University, 751 21 Uppsala, Sweden
c -Department of Chemistry, Box 523, Uppsala University, 751 20 Uppsala, Sweden
d -Laboratoire Leon Brillouin, (CEA-CNRS), CEA/Saclay, 91191 Gif-sur-Yvette, France
e -Department of Physics and Astronomy, Box 516, Uppsala University, 751 20 Uppsala, Sweden
f -Neutron Physics Institute, Rez, Czech Republic

**Abstract.**
Solid solutions of $(1-x)LaFeO_3$-$(x)PbTiO_3$ ($0<x<1$) have been prepared by conventional solid-state reaction. These complex perovskites have been studied by means of X-ray (XRPD) and neutron powder (NPD) diffraction, complemented with dielectric, magnetic, heat capacity and Mössbauer measurements. Complete solubility in the perovskite series was demonstrated. The NPD and XRPD patterns were successfully refined as orthorhombic ($x \leq 0.7$) and tetragonal ($x \geq 0.8$). A composition-driven phase transformation occurs within the interval $0.7<x<0.8$. The samples with $x<0.5$ showed evidence of long-range magnetic ordering with an G-type antiferromagnetic arrangement of the magnetic moments of the $Fe^{3+}$ cations in the B-site with propagation vector **k** = (0,0,0). Based on the obtained experimental data, a combined structural and magnetic phase diagram has been constructed. The factors governing the structural, dielectric and magnetic properties of $(1-x) LaFeO_3 – (x)PbTiO_3$ solid solutions are discussed, as well as their possible multiferroicity.

* Corresponding author. Tel.: +46-(0)18-471 72 33; fax: +46-(0)18-471 32 70
*E-mail address:* roland.mathieu@angstrom.uu.se (R. Mathieu).

## 1. Introduction

Multiferroic materials (MF) exhibiting ferromagnetism (FM) and ferroelectricity (FE) simultaneously have generated a great deal of technological and fundamental interest during the last years, due to potential applications in spintronics and information storage and sensor technology [1-3]. However, the number of single phase materials that exhibit coexistence of spin and dipole ordering at room temperature is quite limited [4,5]. The scarcity of such MF materials is related to the fact that transition metal d electrons, which are essential for the presence of a magnetic moment, reduce the polar lattice distortion which is required for ferroelectric behavior [4-6]. Thus, an additional structural or electronic driving force is required for ferroelectric and magnetic ordering to coexist [7-9]. Although rare, there are compounds that exhibit magnetic order, ferroelectricity and magnetoelectric coupling. However, when these requirements are met, it often occurs well below room temperature and with much weaker coupling than the magnitude of the product of the dielectric and magnetic susceptibilities suggests.

A potential source of MF materials is complex metal oxides that crystallize in perovskite structure [10-11]. A number of compounds with perovskite type of structure has been found where both long range ordering of the spin configuration (ferro/ferri/or antiferromagnetic) and long range ordering of dipole moments (ferro or antiferroelectric) are observed [12,13]. Coexisting FE and magnetic order has e.g. been found in $BiFeO_3$ and $BiMnO_3$ [14-16]. For these materials, the FE is induced by Bi in the A-sublattice, while magnetic order is driven by the 3d cation in the B-sublattice. One promising direction in the search for MF materials is the exploration of solid solution materials combining *A*-site lone pair ferroelectricity with *B*-site magnetic ordering.

In ref. [17] it was proposed that the maximum value of the square of the linear ME coupling coefficient in a material is directly proportional to the product of the electric and magnetic susceptibilities of the material. Improvement of dielectric and magnetic properties of MF perovskites may be reached in solid solutions of materials previously known for their excellent FE or magnetic properties. In this paper, we report results from comprehensive structural, magnetic and dielectric studies of $(1-x)LaFeO_3-(x)PbTiO_3$ (LFPTO) solid solutions. So far, only limited information is available on this system and many questions are still open [18]. The goals of the present study are to extend previous work on the LFPTO system to the full concentration range, to identify more definitely the spin and dipole order and their concentration and temperature evolution and to achieve a magnetoelectric material using the conversion of ferroelectric $PbTiO_3$ (PTO) to a compound with MF properties.

$PbTiO_3$ is a well-known piezoelectric with a strongly distorted tetragonal perovskite structure and a FE phase transition at ~ 760 K [19, 20]. It is well understood that Ti (3d)–O(2p) hybridization helps in stabilizing the FE distortion in PTO [21]. The FE polarization mainly results from displacement of $Ti^{4+}$ cations with respect to the oxygen cage and hybridization of the lowest energy level state with O(2p) states seems to be a requirement for stabilization of the FE state. A reduction in lattice distortion after doping with a transition metal element at the Ti site is expected. However, PTO is known for its very large lattice distortion (c/a =1.064). It was therefore assumed that the lattice could sustain its tetragonal structure even after substituting a transition metal element like Fe at the Ti site, thus making it possible to induce magnetism in PTO without fatally disturbing its FE behaviour [22]. Also, it has recently been shown that doping moderate amounts of V or Cr at the *B*-site induces magnetic properties in PTO without destroying FE [23]. Curiously, undoped nanocrystalline PTO has been reported to exhibit weak room temperature FM which probably is connected with oxygen vacancies [24].

Among the rare-earth perovskite orthoferrites $LaFeO_3$ (LFO) is a very well-known canted antiferromagnetic (AFM) insulator with an orthorhombically distorted perovskite structure (see containing only trivalent iron and exhibiting a high value of the Néel temperature ($T_N$ ~ 740 K) [25, 26]. Although a number of papers, related to the structural and magnetic properties of LFO, has been reported, only recently presented results in ref. [27] showed MF behaviour in LFO.

Besides that, LFO is ferroelastic at room temperature [28, 29] adding extra functionality to the MF properties. The ferroelastic effect in LFO is characterized by a spontaneous lattice strain of about $2.4 \times 10^{-4}$ [29]. The resulting ferroelastic transformation is caused by individual atomic displacements of La, $O_1$ and $O_2$ with Fe remaining invariant at the inversion center. LFO is a coupled ferroelastic antiferromagnet [28] and the spin configuration is dominantly AFM, with the spin direction essentially along the b-axis, however, with a small FM component along the c-axis. Ferroelastic interchange of the *a* and *b* parameters must hence rotate the AFM spin direction through 90°, but will not affect the weak FM. In this context, it is important to note that the crystallographic origin of the ferroelastic behaviour in this compound is fundamentally different from the well-known ferroelasticity of $Pb(Ti,Zr)O_3$ perovskites [30].

## 2. Motivation

LFPTO solid solutions were studied many years ago because of their extensive usage in various technological applications [31]. The structure-property relationships are of particular interest for substitution of La by Pb in LFO since the $6s^2$ lone electron pair of $Pb^{2+}$ should strongly influence the electron configuration around the La cation. As compared to the pure LFO and PTO phases, the literature on preparation and characterization of LFPTO is relatively scarce [30-32]. Investigations on compositions rich in PTO indicated that LFPTO is relatively difficult to work with, because there is a strong tendency to form significant amounts of different non-perovskite phases, which complicate the analysis [31]. Also, the reported XRPD and dielectric measurements suggested a complex phase diagram with two pseudo monoclinic phases, but these preliminary structural data were of low accuracy and the space groups of the possible compositional polymorphs were not defined. Later, in ref. [32], magnetic measurements on the samples prepared in ref. [31] indicated a strong relation between the structural distortion and the magnetic ordering. Recently in ref. [18], a new MF candidate, $Pb_{0.8}La_{0.2}Ti_{0.8}Fe_{0.2}O_3$ (i.e. LFPTO with x=0.80), was proposed. It was concluded that this composition shows coexistence of FE and FM ordering at room temperature. Indeed, in the earlier magnetic studies, solid solutions containing only 20% [18] or 30 % [32] LFO were found to order magnetically at high temperature, near 650 K. We show in the following that solutions with such low LFO content do not order magnetically. We suspect that the reported magnetic order at high temperature in these compounds is related to minor amounts of $LaFe_{12}O_{19}$ impurity phases, which order ferromagnetically just below 700 K [33].

## 3. Experimental procedure

### 3.1. Sample preparation
Ceramic samples of LFPTO solid solutions are prepared using a conventional solid state reaction route. Stoichiometric amounts of binary oxides $La_2O_3$, PbO, $Fe_2O_3$ and $TiO_2$ are first weighed in stoichiometric proportion using a high precision electronic balance, thoroughly mixed and ball-milled for several hours and then pressed in the form of cylindrical pellets with a diameter 11 mm and a thickness of 5 mm. All reagents had a purity of 99.9% or better. $La_2O_3$ was dried at 900° C immediately prior to weighing. The mixtures were placed in a platinum crucible covered with a lid. The Pt crucible was then put in an alumina crucible that was sealed to an alumina lid with $Al_2O_3$ cements. Such a double-crucible setting was used to prevent the volatilization of PbO. The sample preparation included several stages of calcinations, which efficiently suppressed the formation of undesirable impurity phases. These disks were sintered at different temperatures 1000, 1100 and 1200 °C for 48 h in lead rich environment in order to reduce the weight loss due to the volatility of Pb. The series of grinding and sintering procedures were performed until the XRD pattern showed the expected spectra without impurity lines. After the last ball-milling the final sintering was made at 1250° C for 48 h. The mixture was weighed before and after heat treatment to determine possible Pb loss due to evaporation. In all cases, the weight difference was negligible (<0.01%). Possible parasitic phases can be leached in 10% diluted $HNO_3$.

### 3.2. X-ray powder diffraction

The purity of the powder sample was checked from X-ray powder diffraction (XRPD) patterns obtained with a D-5000 diffractometer using Cu $K_\alpha$ radiation. The ceramic samples of LFPTO were crushed into powder in an agate mortar and suspended in ethanol. A Si substrate was covered with several drops of the resulting suspension, leaving randomly oriented crystallites after drying. The XRPD data for Rietveld analysis were collected at room temperature on a Bruker D8 Advance diffractometer (Ge monochromatized Cu $K_{\alpha 1}$ radiation, Bragg-Brentano geometry, DIFFRACT plus software) in the $2\theta$ range 10-152° with a step size of 0.02° (counting time was 15 s per step). The slit system was selected to ensure that the X-ray beam was completely within the sample for all $2\theta$ angles.

### 3.3. Chemical composition

The chemical composition of the prepared ceramic LFPTO samples was analyzed by energy-dispersive spectroscopy (EDS) using a JEOL 840A scanning electron microscope and INCA 4.07 (Oxford Instruments) software. The analyses performed on several particles showed that the concentration ratios of La/Pb and Fe/Ti are the stoichiometric ones within the instrumental resolution (0.05).

### 3.4. Second harmonic generation (SHG) measurements.

The materials were characterized by SHG measurements in reflection geometry, using a pulsed Nd:YAG laser ($\lambda$= 1.064μm). The SHG signal $I_{2\omega}$ was measured from the polycrystalline samples relative to $\alpha$-quartz standard at room temperature in the Q-switching mode with a repetition rate of 4 Hz. To make relevant comparisons of LFPTO microcrystalline powders and $\alpha$-quartz standard were sieved into the same particle size range because the SHG efficiency has been shown to strongly depend on particle size [34].

### 3.5. Magnetic and dielectric measurements

The magnetization experiments were performed on an MPMSXL SQUID magnetometer and PPMS6000 with VSM option, both from Quantum Design Inc. The temperature dependence of the magnetization was recorded as a function of temperature in magnetic fields of H= 20 Oe (MPMS) and 1000 or 2000 Oe (depending on the sample (PPMS)) using zero-field-cooled (ZFC), field-cooled (FC), and thermo-remanent (TRM) protocols. Magnetic hysteresis loops were recorded at low temperatures (T=10 K).
Dielectric properties of LFPTO ceramic samples were measured using ceramic disks (0.3mm thick) with silver electrodes fired on the both sides. The dielectric constant and loss tangent were derived from an impedance analyzer Agilent 4284A interfaced with a temperature chamber at different frequencies (ranging from 100 Hz to 1MHz). To determine $T_c$, capacitance measurements were made as a function of temperature in an automated temperature controlled furnace interfaced with a computer for data acquisition.

### 3.6. Mössbauer measurements

The used transmission Mössbauer spectrometer was of constant acceleration type, using 512 or 1024 cells for storing the unfolded data. The source, $^{57}$CoRh, was always held at room temperature. The absorbing material, typically around 10 – 15 mg/cm$^2$, was crushed and mixed with a suitable amount ($\approx$ 50 mg/cm$^2$) of boron nitride and spread evenly over the absorber disc, diameter 13 mm. The amount of material was chosen to give optimal thicknesses, resulting in maximum of signal-to-noise ratio [35]. Low temperature measurements, down to 78 K, were done using a flow cryostat of Oxford design, using liquid nitrogen as cooling liquid. The folded spectra (256 or 512 cells) covering a velocity span of ± 12 mm/s or less, were least-square fitted using the "Recoil" program [36]. The center shift, CS, being the sum of the true isomer shift and the second order Doppler shift, is given relative to metallic iron at room temperature. In the paramagnetic case, the magnitude of the quadrupole splitting, |QS|, is given as the peak separation in the symmetrical (no texture assumed) doublet. The quadrupole shift ε, in the used dominant magnetic interaction model is defined as: $\varepsilon = ((v_6-v_5)-(v_2-v_1))/4$, where $v_1, v_2, \ldots, v_6$ are the Lorentzian peak positions with increasing velocity in the fitted sextet. In this case $\varepsilon$ is also reated to nuclear and crystallographic parameters through:

$$\varepsilon = \frac{eQV_{zz}}{4} \cdot \frac{3\cos^2\theta - 1 + \eta \sin^2\theta \cos 2\phi}{2}$$

The quadrupole splitting, QS, is given by:

$$QS = \frac{eQV_{zz}}{2} \cdot \sqrt{1 + \eta^2/3}$$

In these expressions θ and φ are the polar and azimuthal angles for the magnetic field in the principal axes system of the electric field gradient at the Mössbauer atom position. Moreover, the thin absorber approximation has been used, thus the intensities for the peaks for increasing velocities in a sextet are proportional to 3:2:1:1:2:3. The hyperfine field, H, is given in Tesla. The intensity (given in %) is the area of a sextet (or doublet for a non-magnetic pattern) below the base-line compared to the total absorbed area for the Mössbauer pattern.

### 3.7. Specific heat measurements
Specific heat measurements were performed using a relaxation method between 300 K and 360 K on the PPMS6000 system. Heat capacity data in the temperature range 200-800 K were obtained by differential scanning calorimetry using a DSCQ1000 from TA Instruments.

### 3.8. Neutron powder diffraction
Because the neutron scattering lengths of La, Pb, Fe and Ti are different, the chemical composition of the *A*- and *B*-site cations can be estimated by neutron powder diffraction (NPD) with good precision ($b_{La}$ = 8.24, $b_{Pb}$= 9.405 , $b_{Fe}$ = 9.45, $b_{Ti}$ =-3.37 fm), particularly for Fe and Ti because of opposite signs of their scattering factors. The neutron scattering length of oxygen ($b_O$ = 5.805 fm) is comparable to those of the heavy atoms and NPD provides accurate information on its position and stoichiometry.
Registration of NPD patterns versus temperature was performed at LLB (Saclay, France) on the high-resolution neutron powder diffractometer 3T2 ($\lambda$=1.225 Å) at 2 K and 295 K. The powdered samples were inserted in a cylindrical vanadium container.
The powder sample of LFPTO with x=0.1 was studied at 10 K, 295 K and 1000 K on diffractometer MEREDIT (Rez, Czech Republic) with the wavelength of 1.46 Å. Data were collected between 4 and 148° in *2θ* with a step length of 0.08°.
The NPD experimental diffraction patterns were analyzed with the Rietveld profile method using the FULLPROF program [37]. The diffraction peaks were described by a pseudo-Voigt profile function, with a Lorentzian contribution to the Gaussian peak shape. A peak asymmetry correction was made for angles below 35° (2θ). Background intensities were estimated by interpolating between up to 40 selected points (low temperature NPD experimental data) or described by a polynomial with six coefficients. During the refinements the two *A*-type cations (La and Pb) and the octahedrally coordinated metal cations (Fe and Ti) were allowed to vary their occupation on the possible metal sites. The refined atomic coordinates at low and room temperature were used to calculate the magnitude of the tilting angles [38,39]. The IVTON software [40] was employed to characterize the coordination spheres of the A and B-site cations and to obtain bond lengths, volumes of coordination polyhedral and displacements of cations from the centers of the coordination polyhedra.

Since Mössbauer spectroscopy provided strong support for the existence of iron only in the trivalent state, the magnetic structure was refined as an independent phase in which only $Fe^{3+}$ cations were included. The magnetic propagation vector was determined from the peak positions of the magnetic diffraction lines using the K-search software which is included in the FULLPROF refinement package [37]. Magnetic symmetry analysis was then done using the program BASIREPS, also part of FULLPROF [37]. Several magnetic models were tried in the refinement, each employing one additional refinement parameter, corresponding to the magnitude of the magnetic moment. Each structural model was refined to convergence, with the best result selected on the basis of agreement factors and stability of the refinement.

## 4. Results

EDS measurements showed the presence of all the cations in the LFPTO samples. According to the elemental analysis done on 20 different crystallites of each sample, the metal compositions of the LFPTO ceramics are close to the nominal values. Scanning electron micrographs showed a uniform distribution of grains of average size between 1.2-1.5 μm, which was slightly different for different concentrations, but without correlation between grain size and concentration. The oxygen content, determined with iodometric titration, was between 2.98(2) and 3.01(2) for the different samples. All these values are very close to the expected ratios and permit to conclude that the composition of the samples is the nominal one.

As a sensitive and reliable method for establishing the presence or absence of acentric distortions, the SHG technique was used as a test for center of symmetry in the prepared ceramics. These measurements at room temperature gave a negative result for samples with x<0.8. (These samples could still be non-centrosymmetric, but at a level detectable only with sensitivities beyond $10^{-2}$ of quartz [34]). However, a room temperature SHG signal was observed for samples with $x \geq 0.8$ including pure $PbTiO_3$. It should also be noted that an observed strong value of the SHG signal on our $PbTiO_3$ sample is consistent with the SHG efficiency tabulated for this FE in ref. [41]. Thus, these SHG results suggest that LFPTO samples with $0.8 \leq x \leq 1.0$ are non-centrosymmetric.

### 4.1. Magnetic measurements

The temperature dependence of the field-cooled (FC) magnetization of the solutions with $0 \leq x \leq 0.4$ recorded from room temperature to 850 K is shown on Figure 1. The pure (x=0) $LaFeO_3$ system undergoes an antiferromagnetic transition near 760 K, in agreement with earlier data [25,26,42-46] and our differential scanning calorimetry measurements (not shown). The LFPTO compounds with x different from 0 instead show two anomalies, one relatively independent of the $PbTiO_3$ content, near 700 K, and one at a temperature that rapidly decreases with increasing x (marked by arrows in Figure 1). Considering the XRPD results, we have attributed the x-independent 700 K anomaly to minor amounts (< 1%) of $LaFe_{12}O_{19}$ secondary phase (this compound becomes ferromagnetic below 695 K [33]). The x-dependent anomaly is attributed to anfiferromagnetic transition in the main phase (see below).

The magnetic properties of the LFPTO solid solution with $x \geq 0.4$ were also investigated, as seen in Figures 2 and 3. The ZFC and FC curves of the x=0.4 sample coalesce near 330 K, in agreement with the data shown in Figure 1. Additionally, the TRM curve becomes zero above 330K, and a peak is observed in the heat capacity (see inset Figure 1). In the case of x=0.5, a similar behavior is observed, albeit the onset of magnetic ordering is shifted to lower temperatures, near 200K. A relatively weak magnetism is observed for the solution with x=0.6. The antiferromagnetic behavior of the solid solutions is also suggested by the magnetic field-dependence of the magnetization shown in Figure 3. Unfortunately, the hysteresis curves of the $LaFeO_3$-rich solutions include the (ferro)magnetic contribution from the $LaFe_{12}O_{19}$ phase. It is interesting to note that the parent compound (x=0, LaFeO$_3$) displays a large coercivity at low temperatures (see insert of Figure 3), yet becomes very soft magnetically at room-temperature. We thus conclude that the x-dependent features in the M-T curves of the solutions with 0 ≤x ≤0.5 reflect the antiferromagnetic state of the systems, whose transition temperature gradually decreases with x content, to vanish above x=0.5. A very similar observation had been found for BiFeO$_3$-Na$_{0.5}$Bi$_{0.5}$TiO$_3$ solid solution system [47].

## 4.2 Dielectric measurements

Dielectric properties of all sintered LFPTO samples were measured at different frequencies (0.1 kHz<$f$<1 MHz) in the temperature range 300 <$T$<800 K. Representative plots of dielectric constant and loss curves for several samples are shown in Figures 4 and 5, respectively.
For pure PbTiO$_3$, the dielectric peak can be observed around 750 K (without frequency dispersion), which corresponds well to textbook data [48]. Frequency independent peaks in ε versus $T$ were also observed for samples with x=0.8 and 0.9. An important feature to note is the large decrease in $T_c$ of the solid solution by simultaneous doping of La$^{3+}$ at $A$-site and Fe$^{3+}$ at $B$-site in PbTiO$_3$ (710 K for x=0.9 and 440 K for x=0.8). It should also be noted that the $T_c$ of the solid solutions changes very little with altered sintering temperature and atmosphere. At the same time, for all LFPTO samples, the temperature dependence of the dielectric constant depicts a typical relaxor behavior [49,50]: A broad dielectric maximum shifts toward higher temperature with increasing frequency. The structural disorder and compositional fluctuations produced in the arrangement at both cation sublattices lead to microscopic heterogeneity in composition. The dielectric permittivity curves for different frequencies merge at high temperatures for the LFPTO samples, in similarity with the usually observed behavior in Pb-based ferroelectric relaxors [51]. In spite of the apparent weak concentration dependence of the relaxor features (see Figure 4), we believe that the observed relaxor behavior has an intrinsic origin, rather than e.g. a Maxwell-Wagner character associated with defects or grain boundaries. First, our ceramics were found to be very dense (measured density between 93 and 95%). Then we have performed scaling analyses of the temperature-dependent dielectric data, which suggest a Debye contribution to the relaxor behavior (see [52]) for all compositions between 0.3≤x≤0.9, with activation energy ranging with concentration from 1.5 to 1.0 eV.

## 4.3. Mössbauer spectroscopy

Mössbauer spectra recorded at room temperature for *iron poor samples* LFPTO (x ≥ 0.7) could all be well fitted (see Figure 6a) using only one doublet having CS = 0.38(1) mm/s and |QS| = 0.42(7) mm/s. *Iron rich samples* LFPTO with x ≤ 0.2 are magnetically ordered at room temperature and could all be reasonably fitted using mainly one sextet (x ≤ 0.1) or three sextets (x = 0.2), used to describe a limited distribution in hyperfine fields. CS = 0.39(3) for these samples are within uncertainty the same as for the iron-poor samples. In the intermediate region, 0.3 ≤ x ≤ 0.6, both a doublet and sextets show up in the same spectrum, but in different fashions. For x = 0.3 (see Figure 6b) and especially for x = 0.4 the "sextet profile" is quite broad indicating a wide distribution of hyperfine fields. Surprisingly, a *narrow* but weak magnetic sextet shows up in the room temperature spectra for both x = 0.5 and 0.6 (see Figure 7a), having hyperfine fields of 51(1) T at room temperature. To test the influence of annealing on a possible chemical disorder, both samples were annealed at 1200 °C for 2 days. The intensities of the magnetic pattern were after normal annealing 18% and 15% for x = 0.6 and 0.5 respectively. After long time annealing the intensities were 11% and 15% respectively, thus indicating that no major change occured.
Samples with x = 0.5 and 0.6, were also studied at temperatures down to 78 K. For x = 0.6 the intensity of the sextet was the same as at room temperature, 18 %. Both the doublet and the sextet were as sharp as at room temperature. For x = 0.5 the intensity of the doublet decreased at lower temperatures, being 85 %, 79 %, 50 % and 32 % at 295 K, 207 K, 133 K and 78 K respectively. The achieved magnetic pattern was quite broad, but a narrow sextet pattern, similar to that obtained at room temperature, was easily recognized also at low temperature (see Figure 7b).

The obtained Mössbauer parameters are given in Table 1, where also the calculated average hyperfine field per iron atom is included and also presented in Figure 8, except for values for $x = 0.6$ and $0.5$ where the individual fields are shown instead. For these compositions a two phase situation seems to occur. The broad magnetic profiles were fitted with a suitable number of broad sextets. CS and ε values within brackets in Table 1 are constrained values to facilitate in estimating the average hyperfine fields.

Mössbauer spectroscopy investigation showed that the obtained CS- and QS- values in the PM state are typical for trivalent iron in an octahedral environment of oxygen atoms. The center shifts are within uncertainties the same as earlier obtained by us in *e.g.* the perovskites $BiFeO_3$ and $Pb(Fe_{0.6}W_{0.2}Nb_{0.2})O_3$ [53,54]. In some perovskites (*e.g.* $SrFeO_3$) $Fe^{4+}$ are observed as CS-values $\leq 0.10$ mm/s are obtained at room temperature [55]. As all CS-values at room temperature in our case are $\approx 0.4$ mm/s tetravalent iron can be excluded. The hyperfine field values at room temperature for $x = 0.2, 0.1$ and $0.0$ are 45.9 (average values), 51.0 and 52.8 Tesla respectively. Pina et al. [56] got CS = 0.395 mm/s and H = 52.8 T for $LaFeO_3$ ($x = 0.0$) in full agreement with our result above. The high H-values are also very typical for high spin trivalent iron. The *wide* hyperfine field distribution might reflect the amount of iron in the neighbourhood of the individual Mössbauer atoms, giving even a paramagnetic signal when the local iron concentration is low enough. The averaged hyperfine field per iron atom when all iron atoms are taken into account is then $\approx 20$ T for $x = 0.4$ and $\approx 37$ T for $x = 0.3$. The relation between composition x and averaged hyperfine field then indicates that the magnetic ordering temperature ($H_{average} = 0$ T) is close to room temperature for $x \approx 0.45$. However, the *distinct* PM and AFM patterns for $x = 0.6$ and $0.5$ needs another explanation. Maybe these compositions are not really homogenized, containing iron rich and iron poor perovskites in the same sample. This view is also supported by the measurements at low temperature, showing that the iron rich phase has a nearly saturated hyperfine field already at room temperature. Defining the transition PM to AFM composition when intensity of the PM pattern is 50 % again gives $T_N$ around room temperature for $x \approx 0.5$.

## 4.4 X-ray powder diffraction

X-ray powder diffraction analysis of room temperature data showed that the prepared samples formed powders with perovskite structure and complete solubility was observed between the end-members of the series (see Figure 9). The position of the peaks varied systematically, which indicates that LFPTO ceramics form a continuous series of solid solutions over the whole range of compositions. The evolution of the lattice parameters of the LFPTO samples with concentration is presented in Figure 10. For samples with $0 \leq x \leq 0.7$, the patterns indicate an orthorhombic structure (Pnma) and for $x \geq 0.8$ a tetragonal distortion (P4mm) appears.

The XRPD indicated the presence of a very small amount (around 1 %) of $Fe_2O_3$ and $LaFe_{12}O_{19}$ phases as impurities only in the sample with x=0.1. Throughout the series, the Goldschmidt tolerance factor *(t)* [57] increases from 0.954 to 1.019. As the concentration of $PbTiO_3$ increases, the average unit cell parameter $V^{1/3}$ increases and this increase is particularly noticeable in the region of tetragonal symmetry (0.8<x<1.0). In the region of orthorhombic phase ($0 \leq x \leq 0.7$) the *a* parameter decreases up to x=0.3 and slightly increases for 0.4<x<0.8. The same tendency was found for the *b* parameter but in this case a strong increase is evident above x=0.4. For x>0.8, both the *a* and the *b* parameters decrease significantly. An opposite behaviour was found for the *c* parameter, which slightly increases with increasing x up to x=0.8 and remarkably increases within the concentration range 0.8<x<1. This complex behaviour and the irregular changes of the lattice metrics with clearly developed minima and maxima may be connected with the opposite influence of substitution on the *A*- and *B*-sublattices, since the radius of $Fe^{3+}$(0.645 Å) is larger than that of the $Ti^{4+}$(0.605 Å) but at the same time the radius of $Pb^{2+}$ (1.49 Å) is larger than that of the $La^{3+}$ (1.36 Å) for coordination number 12.

The observed change in the lattice parameters is an indication of the formation of solid solutions. As the amount of PbTiO$_3$ increases above 80%, the tetragonal distortion *(c/a)* increases as expected. XRPD patterns of the samples with x=0.8-1.0 confirm a single phase tetragonal structure at room temperature with splitting of the (100), (110), (200), (210), and (211) peaks. The crystal structure was refined with the Rietveld method. All the samples with x≥0.8 are well fitted with *P4mm* space group using the structural model of PbTiO$_3$ [58]. The crystallographic characteristics of the tetragonal series are given in Table 2 and selected interatomic distances and angles in Table 3. For pure PbTiO$_3$, the *c/a* ratio is reported to be 1.063 [58]; simultaneous substitution of La$^{3+}$ at the *A*-site and Fe$^{3+}$ at the *B*-site reduces the lattice distortion, as expected from its reduced tolerance factor. It was found that the values for the displacement factors on the Pb sites were quite large as shown in Table 2. The reason is probably the presence of some kind of disorder on the Pb sites leading to a large uncertainty in the atom position [59].

In the case of compounds with 0<x<0.8 all reflections in the room temperature XRPD data were compatible with a single-phase orthorhombic perovskite. The orthorhombic unit cell with a ~ √2a$_p$, b ~ √2a$_p$ and c ~ 2a$_p$ which is well-known for the LaFeO$_3$ structure [60] has been successfully applied for the indexing of the XRPD patterns (with a$_p$ ~ 3.9 Å) and the *Pnma* space group was selected for description of the crystal structure of these LPFTO samples. A summary of the atom positions, lattice parameters and reliability factors of the refinement using the space group *Pnma* is provided in Table 2, and the corresponding difference profile is shown in Figure 9. Selected interatomic distances and bond angles are listed in Table 3.

The lattice parameters for pure LFO and PTO are in good agreement with those reported in refs. [58, 59-62]. There is, however, some disagreement with the cell parameters of the ceramic samples of LFO reported in ref. [63]. This difference in lattice parameters could be explained by different synthesis methods.

## 4.5. Neutron powder diffraction

The crystal structure of the LFPTO samples was also independently determined using NPD data measured above the magnetic ordering temperatures. To assign the proper space group for the description of the LFPTO structure several centrosymmetric orthorhombic space groups were initially considered including the list of different structures where both ferroelectric cation displacements and tilting of octahedral units are present (see [64]). Rietveld refinements were carried out in all space groups, but a clearly superior fit was obtained using the space group *Pnma* within a structural model for pure LFO [60]. This model describes a random distribution of Fe and Ti in the B-site. The distribution of the La and Pb cations at the *A*-site was found fully disordered (see Table 4). During our refinement the relative concentrations of the A and B-site cations were constrained to the overall bulk stoichiometry but the fractions of La/Pb and Fe/Ti cations were allowed to vary. The oxygen concentration was refined to a value close to 1.00 (in the frames of 2 standard deviations). No extra peaks or additional splitting of the main reflections were observed, which would indicate the need for symmetry lower than ortorhombic. After the final refinements we obtained reasonable reliability factors, included in Table 4 together with the final unit cell, atomic parameters and selected bond distances (see Table 5). The orthorhombic and tetragonal crystal structures of LFPTO in polyhedral representation is shown in figure 1. All distances are in reasonable agreement with the expected values from the ionic radii sums for La$^{3+}$/Pb$^{2+}$ cations (coordination number is between 9 and 12) and Fe$^{3+}$/Ti$^{4+}$ cations (coordination number is 6). From individual A-O and B-O distances the valences of cations were calculated at room temperature following the BVS method proposed in [65,66]. This phenomenological model can help to give estimates of the actual valences on the cations of a given structure, by means of an empirical relationship between the observed bond lengths and the valence of the connected ions. The results of the calculations (see Table 6) clearly indicate the validity of the proposed structural models. The cations exhibit valences, which are only slightly different from those expected in this compound. The observed, calculated and difference profiles of LFPTO (x=0.2) at different temperatures are shown in Figure 11.

## 4.6. Magnetic structure determination

Our magnetic measurements indicate that the samples with x≤0.4 become antiferromagnetic already above room temperature. Thus, magnetic reflections are expected in the diffractograms of those samples at low temperatures. This is exemplified in Figure 12 showing neutron powder diffractograms for LFPTO with x=0.10 at 1000, 300 and 10 K. Additional maxima, expected for an antiferromagnetic state below $T_N$ are clearly visible, indicating that a long-range magnetically ordered state is achieved. Moreover, the intensity of these peaks decreases as the temperature increases and become negligible above $T_N$, confirming them to be magnetic in origin. These peaks can be indexed on the basis of a magnetic cell, which is the same as the chemical cell, leading to a propagation vector k = (0,0,0). By using the BASIREPS program included in the FULLPROF suite we tried to do representational analysis of the space group *Pnma* using the propagation vector k = (0, 0, 0) on the magnetic Fe atoms. After checking all four possible irreducible representations for the $Fe^{3+}$ cations ($\Gamma_1$, $\Gamma_3$, $\Gamma_5$ and $\Gamma_7$) (see Table 7) the magnetic structure that shows the best agreement with the experimental data corresponds to the representation $\Gamma_7$ (so called G-type of magnetic structure where inter- and intralayer spin coupling are antiparallel). This mode specifies the magnetic moments mainly along the b axis slightly tilted in the c axis direction according to the tilting of the Fe(Ti)O$_6$ octahedra (so-called $G_yF_z$ mode). By symmetry, all three directions of the magnetic moments are allowed, but we just refined the components $\mu_y$ and $\mu_z$. An attempt to fix the spins along the c-axis resulted in a significant worsening of $R_{mag}$ and an attempt to refine the component $\mu_x$ leads to unstable refinement without any improvement of the R-factors. The magnitude of the magnetic moment at 2 K of the Fe cations for the samples with x=0.1 and 0.2 was found to be in good agreement with the value for pure LaFeO$_3$ (4.6 (2) $\mu_B$/Fe) and thus somewhat smaller than the expected moment for $Fe^{3+}$ ions (5.0 $\mu_B$). For larger x values the moment per Fe decreases further. The total magnetic moment of the unit cell is zero in the ideal antiferromagnetic structure illustrated in Figure 13.

**Discussion**

Many perovskite-type materials show common structural instabilities: cation displacements which drive ferroelectric properties and tilts of the BO$_6$ octahedra which play a key role regarding the magnetic properties (e.g. the Fe-O-Fe bond angle is a critical parameter in the magnetic superexchange). These instabilities are known to be very sensitive to external perturbations such as temperature and pressure. Another way of altering these structural instabilities, and thus the related physical properties is "crystallochemical strain" associated with cation substitutions. In the case of our compounds, stabilizing structural distortions are related to small shifts of the La/Pb and Fe/Ti sublattices.

To gain a full understanding of structural data for the LFPTO samples, the IVTON software [40] was employed to perform a polyhedral analysis of the structure and characterize the coordination spheres of the *A* and *B* cations, and to obtain bond lengths and displacements of the cations from the centers of the coordination polyhedra. Tilt angles of the BO$_6$ octahedra were calculated from bond angles following refs. [38,39]. The obtained results are given in Table 8. The first observation we can make is that the *A*-site cations have shifted significantly away from the centre of its coordination polyhedron. Significant variation of the A-O distances was found (see Tables 5 and 6). The *B*-site cations have also moved away from the octahedral centers but these shifts are significantly smaller than the *A*-cation shifts. As a consequence, the variation in B-O distances is quite small (see Tables 5 and 6).

Octahedral tilting in orthorhombic perovskites results in a reduction of the lattice parameters perpendicular to the tilt axis saving the parameter parallel to this axis unchanged. Therefore, the tilting angles can be roughly estimated using the related unit cell metrics. For orthorhombic LFPTO compositions the observed variation of the lattice parameters suggests that the tilting reaches its maximum value in the middle of the concentration range. This is not an unexpected feature, given that the compositions with x ≈ 0.4-0.6 host the maximum number of cations of different size and charge at both *A*- and *B*-sites. The octahedral tilting angles derived directly from the structure parameters become larger with $PbTiO_3$ doping. The tilting results in reduction of the coordination of an *A*-site cation ($CA_N$) (12 in the undistorted cubic structure). In most orthorhombic perovskites, $CA_N$ may vary between 8 and 12, depending on the tilt angle and ratio between the sizes of *A*- and *B*-site cations. For LFPTO the value of the *A*-site coordination polyhedron $CA_N$ increases with *x*, ranging between 9 and 12. Typically, eight short A-O bonds are in the first coordination sphere and four longer A-O bonds give rise to the second coordination sphere.

The spontaneous orthorhombic strain, defined as *s = (a-c)/(a+c)*, is a measure of the octahedral distortion and tilt angles around [101], [010] and [111]. This parameter progressively increases as $PbTiO_3$ is added up to x=0.5, as shown in Table 8. Analysis of (La, Pb)-$O_{12}$ cages reveals a crucial difference between the bonding behavior of the two *A*-site cations. Pb displacements create three or four very short Pb-O bonds. Thus, the difference in ionic sizes of Pb and La gives rise to different A-O bonding motifs. The LFPTO perovskite series examined in this study is unusual, in as much as it involves two cationic substitutions with essentially opposite structural effects. At the *A*-site, $La^{3+}$ is substituted by the significantly larger $Pb^{2+}$ (1.36 and 1.49 Å, respectively), whereas at the *B*-site, $Fe^{3+}$ (0.645 Å) is replaced by the smaller $Ti^{4+}$ (0.605 Å). This replacement results in an unexpected concentration dependence of lattice metrics (see Figure 10 a-c). Contrary to the linear dependence on composition observed for Pb-based perovskites, LFPTO solid solutions exhibit strongly nonlinear and nonmonotonic dependences of the lattice parameters for orthorhombic compounds ( 0≤x≤0.7). The tolerance factor t, which is a quantitative measure of the structural perfection of perovskites, increases with x (see Table 2). The "ideal" average effective *A*-site cation radius for $A FeO_3$ perovskite is 1.492, i.e. larger than the ionic radius of $La^{3+}$. It is clear that in the case of LFPTO, the mixture of $La^{3+}$ and $Pb^{2+}$ on the *A*-site could produce an average radius suitable for ideal perovskite (with t closest to 1) only for large value of doping x. Since there are trivalent and divalent cations whose ionic radii are not identical or close to each other (the size difference is about 9%), *A*-site disorder will always exist in LFPTO. As for the *B*-sublattice, $Ti^{4+}$ is smaller than $Fe^{3+}$ (the size difference is about 6%) and mixed $Ti^{4+}/Fe^{3+}$ cations are located at the center of the *B*-octahedra, which build the framework of the perovskite and affects the lattice metrics more critically. $PbTiO_3$ doping makes the perovskite lattice less distorted (see Figure 14 and the values of t for different x in Table 2). In the case of LFPTO a complex combination of opposite influence of the *A*- and *B*-sublattices may be at the origin of the nonmonotonic dependence of the lattice parameters and explain the steep increase seen in the range 0.5≤ x ≤0.6. The octahedral (Fe/Ti) framework is an important factor for small x values, whereas the doping in the *A*-sublattice is more important for x>0.4.

The structure of LFPTO is characterized by several different '180° Fe-O-Fe bonds' (see Table 3). The magnetic structure can be explained from AFM superexchange interaction along the Fe-O chains where the Fe-Fe distance is about 3.9 Å. Changes in bond distances and bond angles are known to influence the magnetic properties of systems having indirect exchange interactions. It was found that the Fe-O-Fe angle increases with small concentration of $LaFeO_3$. At these doping levels (x=0, 0.1 and 0.2) the compounds order in an antiferromagnetic structure. The observed variation of $T_N$ with composition can be explained by two governing factors for the magnetic ordering: the dilution of iron cations and the variation of the distance of the magnetic interaction. At higher PTO content no magnetic ordering should occur because the much-diluted iron cations become magnetically isolated below the percolation limit. The increased weak ferromagnetism in the doped samples may be explained using the following scenario: antiferromagnetic spins of $Fe^{3+}$ cations on the B-sites are not fully compensated because of the random distribution of nonmagnetic $Ti^{4+}$ cations, affecting the balance between the antiparallel sublattice magnetizations.

When LaFeO$_3$ is combined with PbTiO$_3$ in solid solution form, Fe$^{3+}$ and Ti$^{4+}$ occupy the B-sites of the lattice at random, which may result in creation of oxygen vacancies to maintain charge neutrality. Simultaneously, *A*-site substitution with Pb$^{2+}$ can cause a distortion in the structure and alter the bond angle of Fe-O-Fe affecting the change in the overall magnetization.

It has been known for some time that the most favorable piezoelectric properties in Pb(Zr,Ti)O$_3$ (PZT) are found at the morphotropic phase boundary (MPB) between the rhombohedral and tetragonal ferroelectric phases [48,49]. PZT solutions are the current materials of choice for piezoelectric applications, due to its high piezoelectric performance at the MPB composition [30]. MPB between the phases with a different symmetry has been found in many Pb-based perovskite solid solutions as a narrow compositional region of multi-structural coexistence with an abrupt change in the crystal structure [30,48,49]. The position of the MPB is strongly influenced by the crystal chemistry of the additive and by the magnitude of the ionic displacement of the "ferroelectrically active" cations on the *B*-site sublattice [30]. Near the boundary there is a delicate microstructural equilibrium that should favor easy local atomic rearrangements which can be achieved with help of new additional "matching" phases. Possibilities of isomorphous substitutions and predicting the solubility limits in perovskite systems including PbTiO$_3$ were studied in details in [67]. In the case of LFPTO a continuous series of solid solutions was derived on the basis of thermodynamic considerations. Since the MPB observed in Pb-based perovskites is located at the transition between two different symmetry structures, it seems intuitive that there could be a relationship between the location of the MPB within a solid solution and the tolerance factor. In the case of LFPTO solid solutions the position of MPB was predicted around 80% PbTiO$_3$ [67,68]. This prediction is in agreement with our observations that the morphotropic phase boundary region (if it exists) in this system covers compositions between 0.7<x<0.8.

## 5. Conclusions

In summary, we have successfully synthesized powder samples of (1-x)LaFeO$_3$-(x)PbTiO$_3$ solid solutions (0<x<1) using a solid state reaction route and conditions for synthesizing single phase samples have been established. The magnetic and dielectric properties of the (1-x) LaFeO$_3$ – (x)PbTiO$_3$ system are schematically summarized in the electronic phase diagram shown in Figure 15. Co-existence of magnetic and certain dielectric order at room temperature is observed for samples in a rather narrow range around x=0.4. In order to clarify the low-temperature dielectric properties and possible magnetoelectric effects in doped perovskites with x<0.8, dielectric spectroscopy experiments up to 1.5 K being performed.

**Acknowledgements**

Financial support from the Swedish Research Council (VR), the Swedish Foundation for International Cooperation in Research and Higher Education (STINT) and the Russian Foundation for Basic Research is gratefully acknowledged.

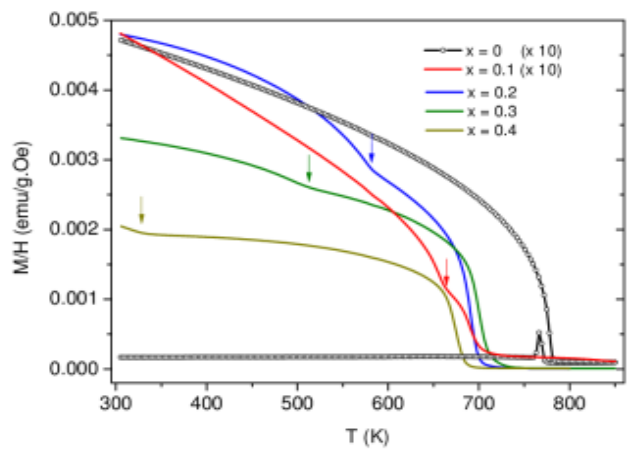

Figure 1

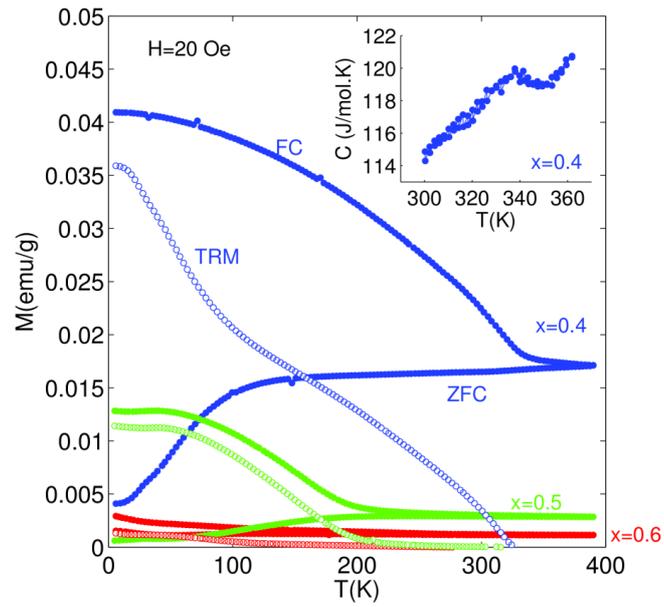

Figure 2

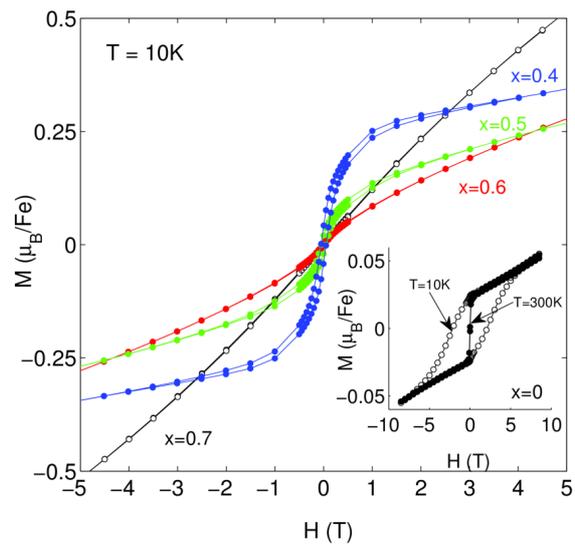

Figure 3

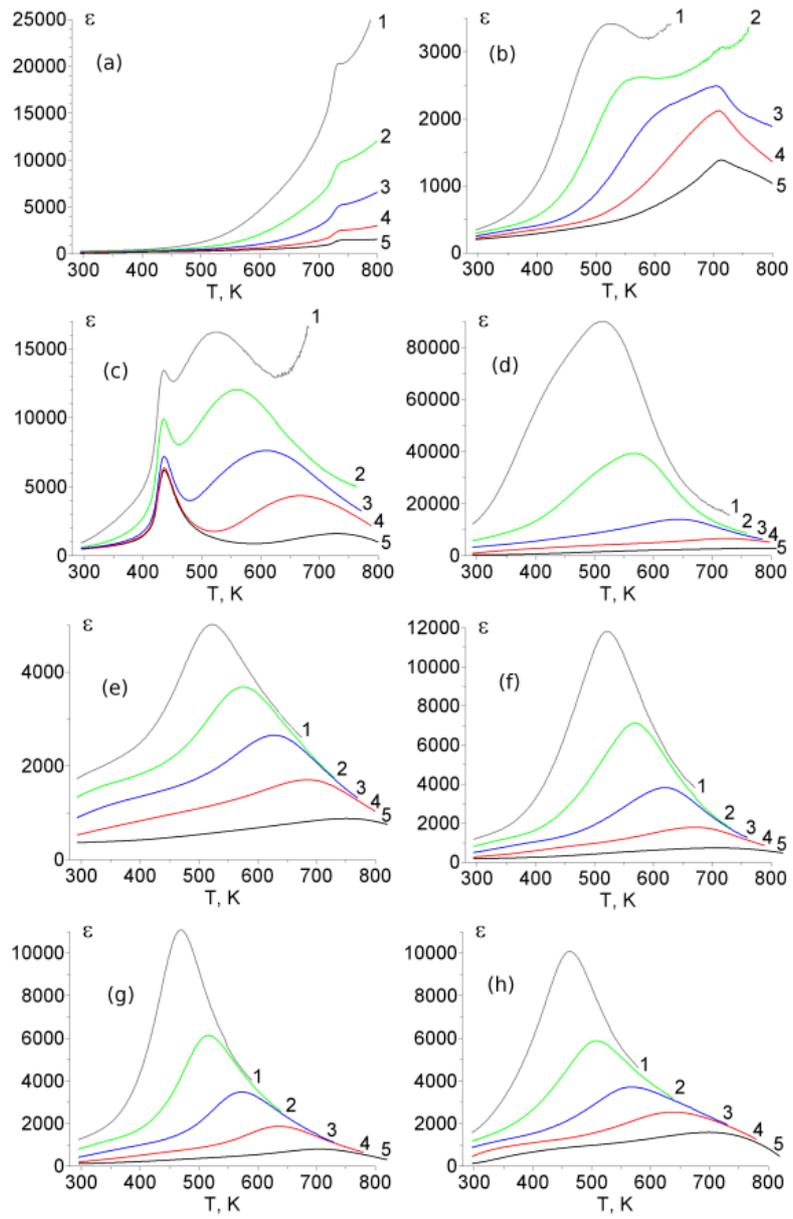

Figure 4

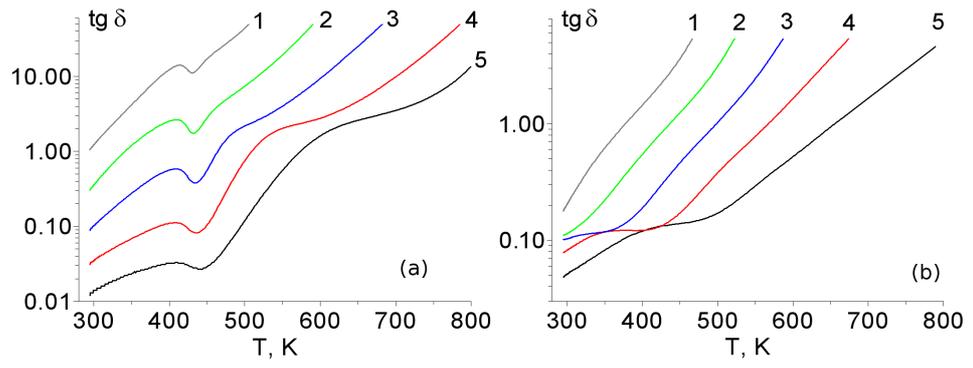

Figure 5

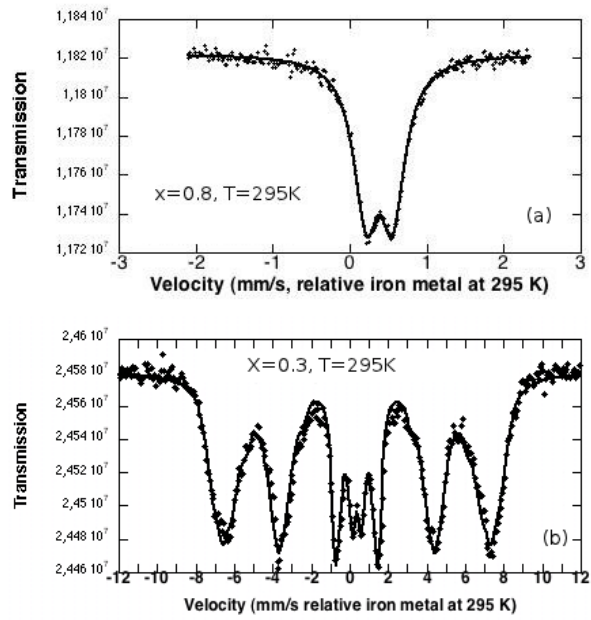

Figure 6

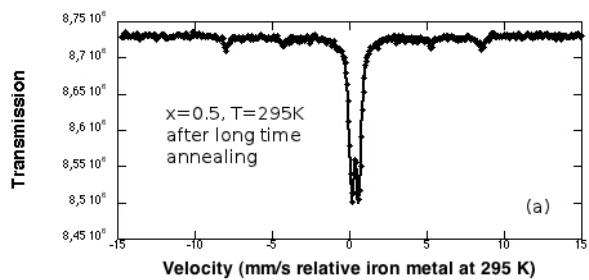
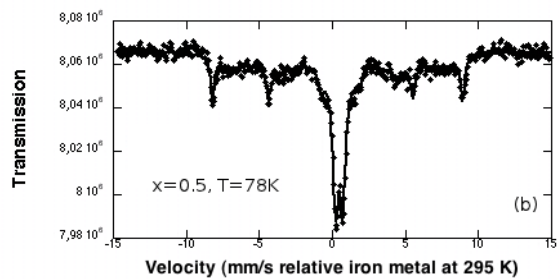

Figure 7

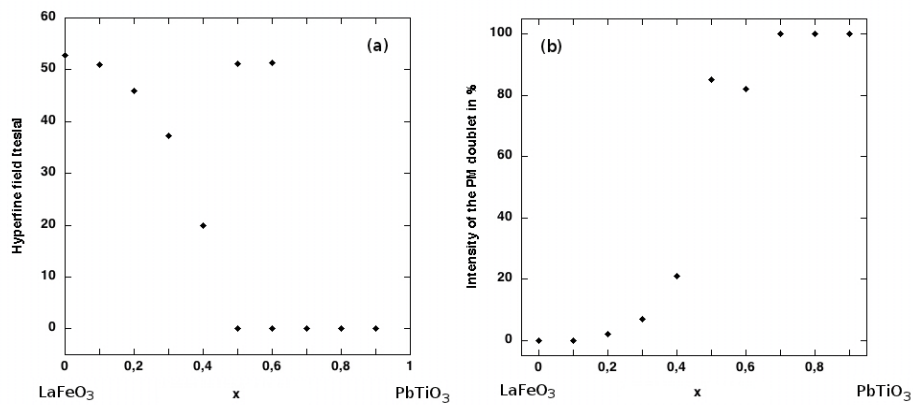

Figure 8

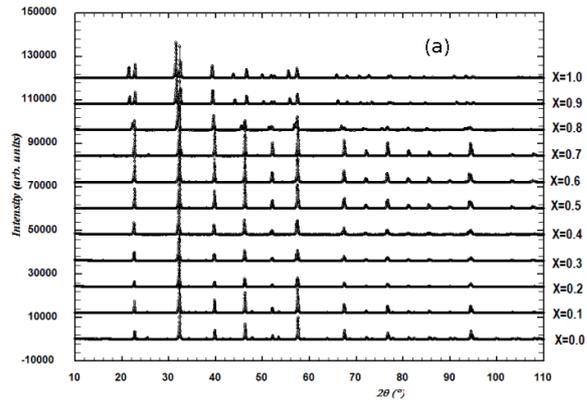

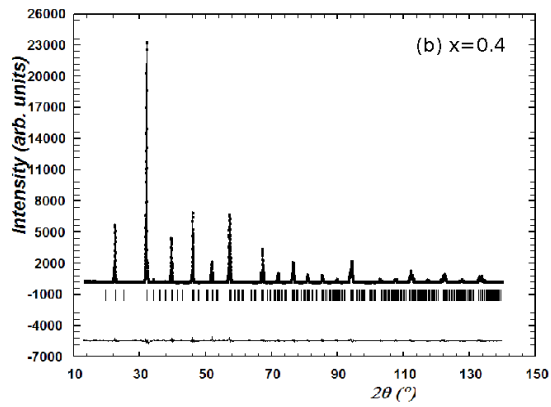

Figure 9

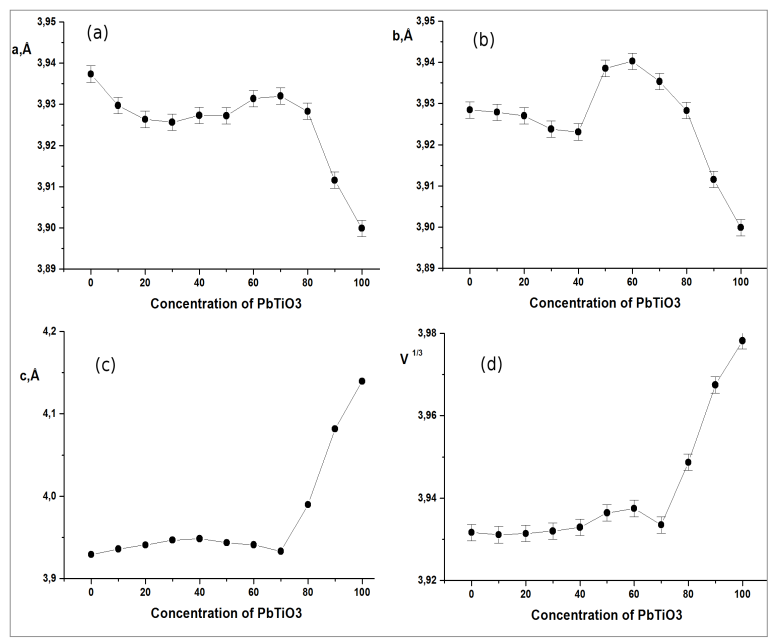

Figure 10

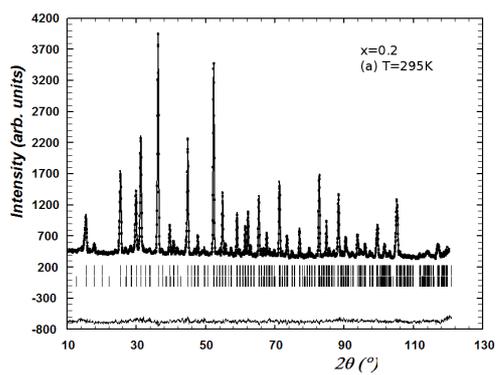

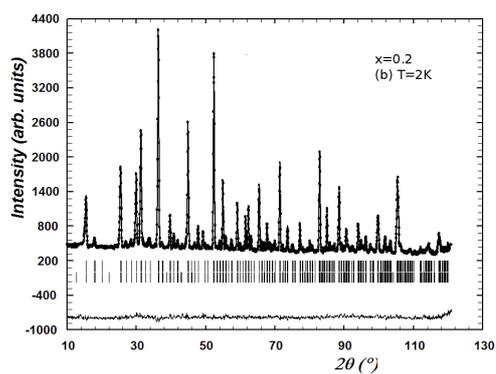

Figure 11

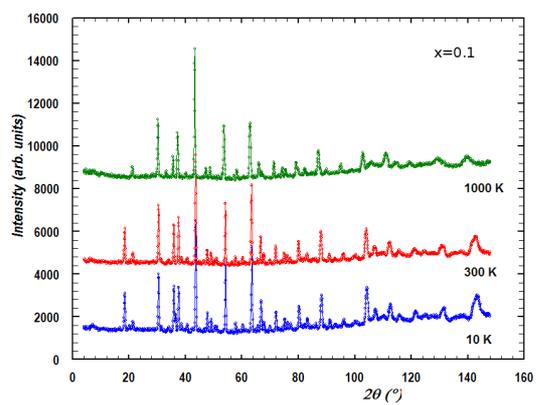

Figure 12

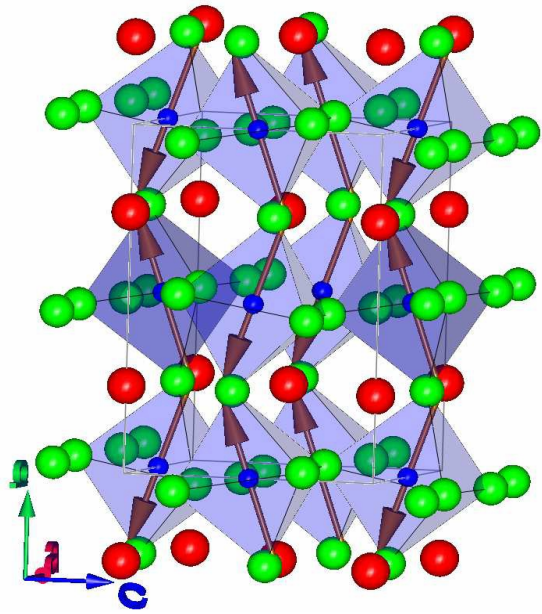

Figure 13

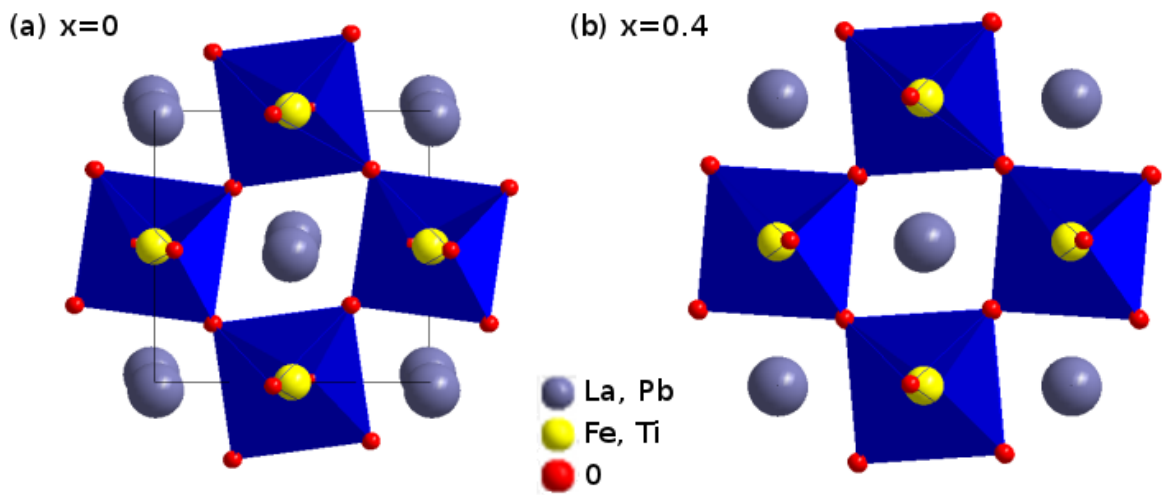

Figure 14

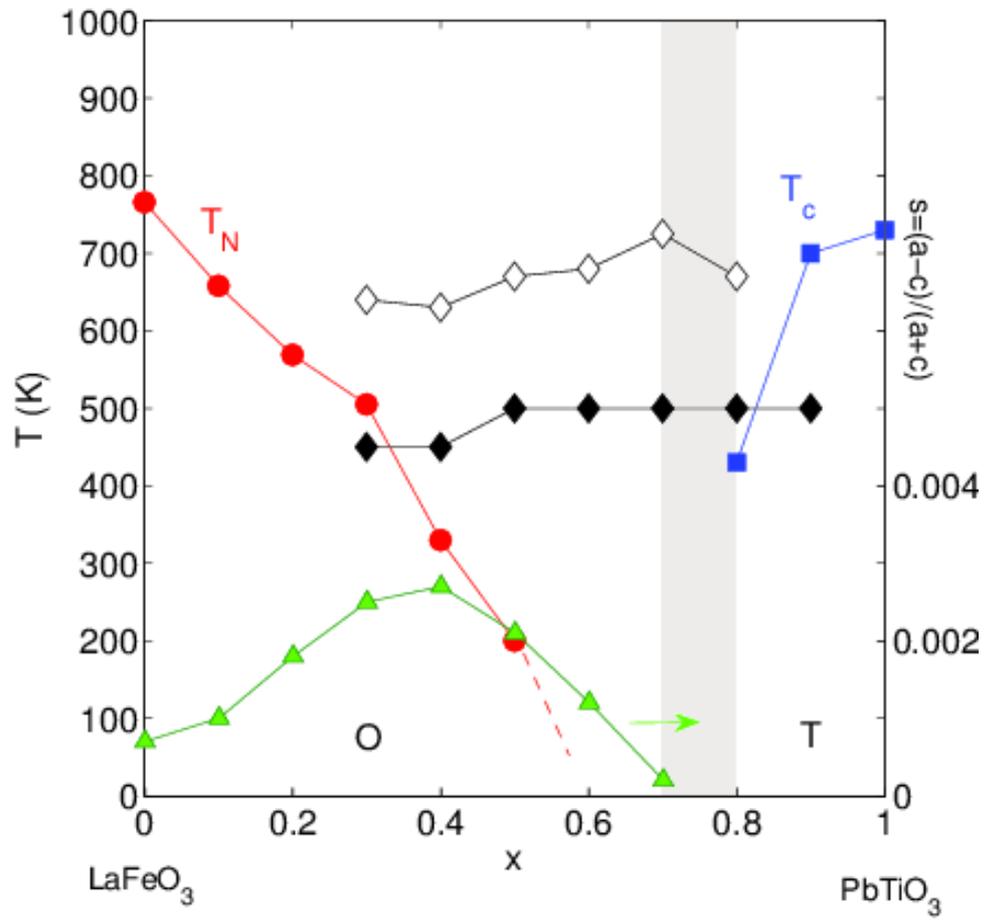

Figure 15

**Table 1** Obtained Mössbauer parameters for LFPTO solid solutions.

| x = | T[K] | CS [mm/s] | |QS|, ε [mm/s] | H[T] | Int.[%] | $H_{average}$[T] |
|---|---|---|---|---|---|---|
| 0 | 295 | 0.36 | -0.04 | 52.8 | 100 | 52.8 |
| 0.1 | 295 | 0.38 | -0.04 | 51.0 | 100 | 51.0 |
| 0.2 | 295 | 0.41 | 0.43 | 0 | 2 | |
| | | 0.37 | -0.03 | 49.2 | 42 | |
| | | 0.38 | -0.04 | 46.4 | 38 | |
| | | 0.38 | 0.10 | 41.7 | 18 | 45.9 |
| 0.3 | 295 | 0.38 | 0.43 | 0 | 7 | |
| | | (0.38) | -0.06 | 45.7 | 19 | |
| | | (0.38) | 0.025 | 41.6 | 62 | |
| | | (0.38) | 0.08 | 23.1 | 12 | 37.3 |
| 0.4 | 295 | 0.38 | 0.44 | 0 | 21 | |
| | | 0.39 | -0.10 | 48.9 | 8 | |
| | | (0.39) | (-0.10) | 26.0 | 60 | |
| | | (0.39) | (.0.10) | 3.9 | 11 | 19.9 |
| 0.4 annealed | 295 | 0.38 | 0.49 | 0 | 29 | |
| | | (0.38) | (-0.10) | 50.8 | 6 | |
| | | (0.38) | (-0.10) | 25.6 | 62 | |
| | | (0.38) | (-0.10) | 6.0 | 4 | 19.2 |
| 0.5 | 295 | 0.40 | 0.45 | 0 | 85 | |
| | | 0.37 | -0.10 | 51.1 | 15 | (7.7) |
| 0.5 annealed | 295 | 0.38 | 0.45 | 0 | 85 | |
| | | 0.38 | -0.10 | 51.2 | 15 | (7.7) |
| 0.5 | 207 | 0.45 | 0.45 | 0 | 79 | |
| | | 0.42 | -0.09 | 52.3 | 21 | (11.0) |
| 0.5 | 133 | 0.48 | 0.45 | 0 | 50 | |
| | | 0.47 | -0.11 | 52.9 | 20 | |
| | | 0.47 | -.12 | 28.2 | 31 | (19.3) |
| 0.5 | 78 | 0.50 | 0.48 | 0 | 32 | |
| | | 0.48 | -0.11 | 53.0 | 18 | |
| | | (0.48) | (-0.11) | 43.0 | 32 | |
| | | (0.48) | (-0.11) | 25.9 | 19 | (28.2) |
| 0.6 | 295 | 0.39 | 0.43 | 0 | 82 | |
| | | 0.38 | -0.10 | 51.4 | 18 | (9.3) |
| 0.6 annealed | 295 | 0.38 | 0.42 | 0 | 89 | |
| | | 0.37 | -0.12 | 51.3 | 11 | (5.7) |
| 0.6 | 78 | 0.51 | 0.44 | 0 | 82 | |
| | | 0.47 | -0.06 | 53.3 | 18 | (9.6) |
| 0.7 | 295 | 0.39 | 0.49 | 0 | 100 | 0 |
| 0.8 | 295 | 0.39 | 0.36 | 0 | 100 | 0 |
| 0.9 | 295 | 0.37 | 0.40 | 0 | 100 | 0 |

**Table 2** Results of the Rietveld refinements of the crystal structure of the LFPTO samples at room temperature using X-ray powder diffraction data. Standard deviations of occupation factors are generally less than 0.02.

| Phase | | x=0 | x=0.1 | x=0.2 | x=0.3 | x=0.4 | x=0.5 | x=0.6 | x=0.7 | x=0.8 | x=0.9 | x=1 |
|---|---|---|---|---|---|---|---|---|---|---|---|---|
| t | | 0.9543 | 0.9607 | 0.9671 | 0.9735 | 0.9799 | 0.9865 | 0.9929 | 0.9994 | 1.006 | 1.0126 | 1.0192 |
| a,Å | | 5.5682(3) | 5.5574(4) | 5.5526(3) | 5.5517(4) | 5.5540(4) | 5.5537(3) | 5.5598(3) | 5.5602(4) | 3.9283(3) | 3.9115(3) | 3.8999(3) |
| b,Å | | 7.8569(5) | 7.8572(5) | 7.8541(4) | 7.8476(5) | 7.8462(4) | 7.8771(5) | 7.8806(4) | 7.8707(5) | 3.9283(3) | 3.9115(3) | 3.8999(3) |
| c,Å | | 5.5566(3) | 5.5652(4) | 5.5731(3) | 5.5805(4) | 5.5838(3) | 5.5768(4) | 5.5733(3) | 5.5622(4) | 3.9899(3) | 4.0675(3) | 4.1396(3) |
| s.g | | *Pnma* | *Pnma* | *Pnma* | *Pnma* | *Pnma* | *Pnma* | *Pnma* | *Pnma* | *P4mm* | *P4mm* | *P4mm* |
| La/Pb (EDS) | | 1.01 | 0.92/0.08 | 0.78/0.22 | 0.72/0.28 | 0.61/0.39 | 0.52/0.48 | 0.39/0.61 | 0.27/0.72 | 0.22/0.78 | 0.13/0.87 | 0.98 |
| Fe/Ti (EDS) | | 0.99 | 0.89/0.11 | 0.81/0.19 | 0.69/0.31 | 0.58/0.42 | 0.49/0.51 | 0.42/0.58 | 0.32/0.68 | 0.19/0.81 | 0.08/0.92 | 1.02 |
| La/Pb | n La/Pb | 1.01/0.99 | 0.88/0.12 | 0.79/0.21 | 0.73/0.27 | 0.63/0.37 | 0.47/0.53 | 0.37/0.63 | 0.28/0.72 | 0.21/0.79 | 0.12/0.88 | 0.99 |
| | x/a | 0.0296(7) | 0.0225(8) | 0.0163(6) | 0.0089(8) | 0.0035(7) | 0.0035(6) | 0.0037(7) | -0.0028(8) | 0 | 0 | 0 |
| | y/b | 1/4 | 1/4 | 1/4 | 1/4 | 1/4 | 1/4 | 1/4 | 1/4 | 0 | 0 | 0 |
| | z/c | -0.0065(11) | -0.0055(9) | -0.0048(9) | -0.0036(12) | -0.0020(9) | 0.0006(10) | 0.0001(9) | -0.0003(11) | 0 | 0 | 0 |
| | Beq(Å$^2$) | 0.83(6) | 0.81(6) | 0.78(5) | 0.75(5) | 0.71(4) | 0.67(4) | 0.84(5) | 0.92(5) | 1.06(4) | 1.13(4) | 1.45(5) |
| Fe/Ti | n Fe/Ti | 0.98/1.02 | 0.89/0.11 | 0.78/0.22 | 0.68/0.32 | 0.59/0.41 | 0.49/0.51 | 0.38/0.62 | 0.27/0.73 | 0.18/0.82 | 0.10/0.90 | 1.01 |
| | x/a | 0 | 0 | 0 | 0 | 0 | 0 | 0 | 0 | 1/2 | 1/2 | 1/2 |
| | y/b | 0 | 0 | 0 | 0 | 0 | 0 | 0 | 0 | 1/2 | 1/2 | 1/2 |
| | z/c | 1/2 | 1/2 | 1/2 | 1/2 | 1/2 | 1/2 | 1/2 | 01/2 | 0.5074 | 0.5218 | 0.5380 |
| | Beq(Å$^2$) | 0.66(3) | 0.64(2) | 0.62(4) | 0.57(3) | 0.54(4) | 0.52(3) | 0.55(3) | 0.54(4) | 0.57(2) | 0.53(2) | 0.59(2) |
| O1 | n O | 1.02 | 0.98 | 1.01 | 0.99 | 1.01 | 0.97 | 0.98 | 0.97 | 1.03 | 0.98 | 1.02 |
| | x/a | 0.4864(11) | 0.4920(9) | 0.4981(8) | 0.5049(10) | 0.5031(11) | 0.5005(9) | 0.5007(8) | 0.5047(10) | 1/2 | 1/2 | 1/2 |
| | y/b | 1/4 | 1/4 | 1/4 | 1/4 | 1/4 | 1/4 | 1/4 | 1/4 | 1/2 | 1/2 | 1/2 |
| | z/c | 0.0714(12) | 0.0618(10) | 0.0568(11) | 0.0561(10) | 0.0451(9) | 0.0279(9) | 0.0180(11) | 0.0062(9) | 0.0581(10) | 0.0812(11) | 0.1110(9) |
| | Beq(Å$^2$) | 0.94(5) | 0.99(6) | 0.96(5) | 1.08(6) | 1.14(5) | 1.19(6) | 1.23(5) | 1.15(6) | 1.23(5) | 1.11(6) | 1.26(6) |
| O2 | n O | 0.98 | 1.02 | 0.97 | 1.03 | 0.97 | 0.98 | 0.97 | 0.99 | 0.98 | 1.01 | 0.97 |
| | x/a | 0.2839(9) | 0.2775(8) | 0.2691(9) | 0.2675(10) | 0.2662(9) | 0.2401(8) | 0.2440(9) | 0.2446(9) | 1/2 | 1/2 | 1/2 |
| | y/b | 0.0389(8) | 0.0362(11) | 0.0325(10) | 0.0303(9) | 0.0284(8) | 0.0347(9) | 0.0308(11) | 0.0114(12) | 0 | 0 | 0 |
| | z/c | 0.7140(9) | 0.7148(10) | 0.7174(9) | 0.7229(8) | 0.7287(8) | 0.7536(9) | 0.7524(11) | 0.7536(10) | 0.5695(8) | 0.5916(9) | 0.6190(9) |
| | Beq(Å$^2$) | 1.18(6) | 1.23(6) | 1.31(5) | 1.28(6) | 0.96(5) | 0.88(6) | 0.92(6) | 1.24(7) | 1.06(6) | 1.19(6) | 1.42(8) |
| Rp | | 4.16 | 4.41 | 4.11 | 4.29 | 4.54 | 4.66 | 4.87 | 4.77 | 5.11 | 5.08 | 4.74 |
| Rwp | | 5.48 | 5.67 | 5.26 | 5.59 | 5.57 | 5.87 | 6.06 | 5.91 | 6.74 | 6.53 | 5.87 |
| Rb | | 4.34 | 4.52 | 4.17 | 4.26 | 4.47 | 4.67 | 4.73 | 4.85 | 5.08 | 4.98 | 4.81 |
| $\chi^2$ | | 1.18 | 1.23 | 1.09 | 1.32 | 1.16 | 1.11 | 1.24 | 1.31 | 1.17 | 1.28 | 1.14 |

**Table 3** Selected bond distances and angles from XRPD powder refinements of LFPTO samples at room temperature.

| Phase | | x=0 | x=0.1 | x=0.2 | x=0.3 | x=0.4 | x=0.5 | x=0.6 | x=0.7 | x=0.8 | x=0.9 | x=1 |
|---|---|---|---|---|---|---|---|---|---|---|---|---|
| La/Pb | O1 | 3.056(5) | 2.972(4) | 2.898(4) | 2.817(4) | 2.792(5) | 2.798(5) | 2.799(4) | 2.739(5) | 2.787(4)x4 | 2.786(4)x4 | 2.796(3)x4 |
| | O1 | 2.580(4) | 2.636(3) | 2.697(4) | 2.774(4) | 2.787(5) | 2.765(5) | 2.765(4) | 2.822(4) | | | |
| | O1 | 3.152(5) | 3.100(4) | 3.078(5) | 3.083(4) | 3.033(5) | 2.947(5) | 2.887(3) | 2.814(5) | | | |
| | O1 | 2.425(3) | 2.475(4) | 2.499(4) | 2.497(5) | 2.551(4) | 2.629(3) | 2.686(4) | 2.749(5) | | | |
| | O2x2 | 2.678(4) | 2.693(3) | 2.699(4) | 2.714(5) | 2.723(4) | 2.550(4) | 2.583(5) | 2.698(3) | 2.609(3)x4 | 2.566(4)x4 | 2.508(4)x4 |
| | O2x2 | 2.440(3) | 2.456(4) | 2.492(3) | 2.504(3) | 2.523(4) | 2.649(4) | 2.655(3) | 2.733(4) | | | |
| | O2x2 | 3.292(5) | 3.233(5) | 3.162(4) | 3.105(4) | 3.057(5) | 2.957(5) | 2.949(4) | 2.818(4) | 3.003(5)x4 | 3.101(5)x4 | 3.220(5)x4 |
| | O2x2 | 2.780(4) | 2.790(/5) | 2.806(4) | 2.824(5) | 2.839(3) | 3.007(5) | 2.974(4) | 2.881(4) | | | |
| Fe/Ti | O1x2 | 2.006(3) | 1.995(4) | 1.989(3) | 1.987(4) | 1.978(4) | 1.975(4) | 1.973(3) | 1.968(3) | 1.793(3) | 1.79283) | 1.768(3) |
| | O2x2 | 2.001(4) | 1.972(3) | 1.941(2) | 1.952(3) | 1.966(3) | 1.964(4) | 1.969(4) | 1.96284) | 2.197(3) | 2.275(3) | 2.372(3) |
| | O2x2 | 2.017(3) | 2.032(4) | 2.047(3) | 2.028(3) | 2.008(4) | 2.010(3) | 1.997(3) | 1.976(3) | 1.979(3)x4 | 1.976(3)x4 | 1.978(3)x4 |
| Fe-O-Fe | | 156.5(2) | 159.9(3) | 161.7(2) | 161.8(3) | 165.3(2) | 170.9(3) | 174.2(3) | 177.5(2) | 180.0 | 180.0 | 180.0 |
| | | 156.4(3) | 158.3(2) | 161.2(3) | 162.9(2) | 164.5(3) | 163.9(2) | 165.8(3) | 174.2/3) | 165.6(2) | 163.5(2) | 153.6(2) |

**Table 4** Summary of structural refinement results of LFPTO samples using NPD data.

| Phase | | x=0.1 | | | x=0.2 | | x=0.3 | | x=0.4 | | x=0.5 | |
|---|---|---|---|---|---|---|---|---|---|---|---|---|
| T,K | | 10 | 295 | 1000 | 2 | 295 | 2 | 295 | 2 | 295 | 2 | 295 |
| a,Å | | 5.5488(2) | 5.5562(2) | 5.5829(3) | 5.5444(2) | 5.5506(3) | 5.5410(2) | 5.5491(3) | 5.5380(2) | 5.5484(3) | 5.5381(2) | 5.5489(3) |
| b,Å | | 7.8405(3) | 7.8561(3) | 7.9254(4) | 7.8402(3) | 7.8510(3) | 7.8344(3) | 7.8449(3) | 7.8250(3) | 7.8402(3) | 7.8251(3) | 7.8694(3) |
| c,Å | | 5.5569(3) | 5.5686(2) | 5.6245(3) | 5.5601(2) | 5.5701(3) | 5.5667(2) | 5.5774(3) | 5.5707(2) | 5.5790(3) | 5.5709(2) | 5.5745(3) |
| s.g | | *Pnma* | *Pnma* | *Pnma* | *Pnma* | *Pnma* | *Pnma* | *Pnma* | *Pnma* | *Pnma* | *Pnma* | *Pnma* |
| La/Pb | La/Pb | 0.91/0.09 (0.02) | | | 0.82/0.18 (0.02) | | 0.71/0.29 (0.02) | | 0.62/0.38 (0.02) | | 0.51/0.49 (0.02) | |
| | x/a | 0.0252(5) | 0.0250(5) | 0.0128(7) | 0.0194(5) | 0.0158(6) | 0.0112(4) | 0.0076(5) | -0.0025(5) | -0.0009(6) | -0.0023(5) | -0.0056(6) |
| | y/b | 1/4 | 1/4 | 1/4 | 1/4 | 1/4 | 1/4 | 1/4 | 1/4 | 1/4 | 1/4 | 1/4 |
| | z/c | 0.0072(7) | 0.0064(8) | 0.0057(8) | 0.0036(7) | 0.0021(6) | 0.0034(6) | 0.0021(7) | -0.0033(8) | -0.0033(6) | -0.0033(8) | -0.0015(7) |
| | B(Å$^2$) | 0.01(5) | 0.46(6) | 2.06(9) | 0.96(5) | 1.39(8) | 0.78(6) | 1.19(7) | 0.79(6) | 1.12(8) | 0.85(7) | 1.02(8) |
| Fe/Ti | Fe/Ti | 0.91/0.09 (0.02) | | | 0.82/0.18 (0.02) | | 0.71/0.29 (0.02) | | 0.62/0.38 (0.02) | | 0.51/0.49 (0.02) | |
| | x/a | 0 | 0 | 0 | 0 | 0 | 0 | 0 | 0 | 0 | 0 | 0 |
| | y/b | 0 | 0 | 0 | 0 | 0 | 0 | 0 | 0 | 0 | 0 | 0 |
| | z/c | 1/2 | 1/2 | 1/2 | 1/2 | 1/2 | 1/2 | 1/2 | 1/2 | 1/2 | 1/2 | 1/2 |
| | B(Å$^2$) | 0.51(3) | 0.61(5) | 1.30(6) | 0.64(4) | 0.91(5) | 0.91(3) | 1.14(4) | 0.69(3) | 1.06(5) | 0.76(5) | 1.31(7) |
| O1 | O | 0.98(2) | | | 1.02(2) | | 0.97(2) | | 1.01(2) | | 1.03(2) | |
| | x/a | 0.4879(5) | 0.4877(6) | 0.4879(8) | 0.4932(9) | 0.4947(8) | 0.4961(7) | 0.4975(8) | 0.5044(7) | 0.5023(8) | 0.5041(7) | 0.5023(8) |
| | y/b | 1/4 | 1/4 | 1/4 | 1/4 | 1/4 | 1/4 | 1/4 | 1/4 | 1/4 | 1/4 | 1/4 |
| | z/c | 0.0699(6) | 0.0713(8) | 0.0668(9) | 0.0667(8) | 0.0639(7) | 0.0615(7) | 0.0581(8) | 0.0573(9) | 0.0526(7) | 0.0579(8) | 0.0453(9) |
| | B(Å$^2$) | 0.19(3) | 0.21(4) | 1.65(6) | 1.11(3) | 1.38(4) | 1.03(3) | 1.36(4) | 0.99(2) | 1.28(3) | 1.09(3) | 0.97(4) |
| O2 | O | 1.02(2) | | | 0.98(2) | | 0.98(2) | | 0.97(2) | | 0.99(2) | |
| | x/a | 0.2809(4) | 0.2826(5) | 0.2605(6) | 0.2753(4) | 0.2737(6) | 0.2692(5) | 0.2659(4) | 0.2616(6) | 0.2595(5) | 0.2615(5) | 0.2585(6) |
| | y/b | 0.0367(5) | 0.0328(6) | 0.0348(7) | 0.0360(5) | 0.0352(4) | 0.0339(4) | 0.0326(5) | 0.0307(6) | 0.0288(5) | 0.0307(6) | 0.0249(5) |
| | z/c | 0.7210(4) | 0.7251(5) | 0.7404(6) | 0.7251(4) | 0.7278(6) | 0.7309(4) | 0.7350(5) | 0.7401(4) | 0.7425(5) | 0.7400(6) | 0.7429(6) |
| | B(Å$^2$) | 0.24(3) | 0.82(4) | 2.12(7) | 1.13(3) | 1.51(4) | 1.12(3) | 1.37(4) | 1.15(4) | 1.42(5) | 1.21(4) | 1.47(5) |
| Rp | | 4.03 | 4.12 | 6.21 | 4.26 | 3.94 | 4.32 | 4.21 | 5.08 | 5.49 | 5.13 | 5.46 |
| Rwp | | 5.07 | 5.21 | 7.65 | 5.63 | 5.29 | 5.70 | 5.51 | 6.22 | 6.47 | 6.17 | 6.39 |
| Rb | | 3.68 | 3.62 | 6.40 | 5.19 | 5.32 | 5.07 | 5.48 | 5.67 | 5.88 | 5.57 | 5.74 |
| Rm | | 2.89 | 3.27 | | 7.16 | 7.31 | 6.44 | 8.13 | 8.24 | 8.47 | 8.36 | 8.61 |
| $\mu_x$ | | 0 | 0 | | 0 | 0 | 0 | 0 | 0 | 0 | 0 | 0 |
| $\mu_y$ | | 4.1(6) | 3.86(7) | | 3.65(5) | 3.26(6) | 3.18(5) | 2.73(6) | 2.51(6) | 2.14(7) | 1.68(6) | |
| $\mu_z$ | | -1.33(8) | -0.90(7) | | -0.96(7) | -0.77(9) | -0.75(8) | -0.58(9) | -0.52(7) | -0.39(8) | -0.26(7) | |
| $\chi^2$ | | 2.11 | 2.29 | 2.01 | 2.08 | 1.87 | 2.21 | 2.06 | 1.96 | 2.18 | 2.09 | 2.17 |

**Table 5** Selected bond distances and angles from NPD powder refinements of LFPTO samples at room temperature.

| Phase | | x=0.1 | | | x=0.2 | | x=0.3 | | x=0.4 | | x=0.5 | |
|---|---|---|---|---|---|---|---|---|---|---|---|---|
| T,K | | 10 | 295 | 1000. | 2 | 295 | 2 | 295 | 2 | 295 | 2 | 295 |
| La/Pb | O1 | 3.013(3) | 3.017(4) | 2.982(5) | 2.944(3) | 2.916(4) | 2.876(3) | 2.855(4) | 2.751(4) | 2.757(5) | 2.736(4) | 2.743(5) |
| | O1 | 2.602(4) | 2.607(4) | 2.662(5) | 2.655(3) | 2.683(4) | 2.712(4) | 2.735(4) | 2.828(3) | 2.824(4) | 2.836(3) | 2.828(4) |
| | O1 | 3.134(5) | 3.152(4) | 3.158(3) | 3.135(4) | 3.131(5) | 3.108(4) | 3.103(5) | 3.086(5) | 3.077(4) | 3.061(4) | 3.048(5) |
| | O1 | 2.438(3) | 2.432(4) | 2.473(6) | 2.433(3) | 2.444(3) | 2.461(4) | 2.475(4) | 2.485(3) | 2.503(3) | 2.511(4) | 2.527(4) |
| | O2 x2 | 2.693(5) | 2.682(4) | 2.620(6) | 2.664(3) | 2.675(4) | 2.664(4) | 2.666(5) | 2.670(3) | 2.684(4) | 2.658(4) | 2.712(5) |
| | O2 x2 | 2.459(3) | 2.491(4) | 2.573(5) | 2.484(5) | 2.496(5) | 2.504(4) | 2.529(4) | 2.528(3) | 2.556(4) | 2.581(3) | 2.589(4) |
| | O2 x2 | 3.235(4) | 3.211(5) | 3.111(6) | 3.175(5) | 3.151(6) | 3.108(6) | 3.073(6) | 3.006(5) | 2.979(6) | 2.972(5) | 2.941(6) |
| | O2 x2 | 2.796(3) | 2.782(4) | 2.933(6) | 2.819(3) | 2.832(4) | 2.851(4) | 2.870(4) | 2.905(3) | 2.908(5) | 2.893(4) | 2.890(4) |
| Fe/Ti | O1 x2 | 1.999(2) | 2.005(3) | 2.018(5) | 1.995(3) | 1.995(3) | 1.989(3) | 1.988(4) | 1.982(3) | 1.983(5) | 1.978(4) | 1.980(4) |
| | O2 x2 | 2.005(2) | 2.026(2) | 2.005(4) | 1.994(3) | 1.999(4) | 1.986(2) | 1.990(3) | 1.987(3) | 1.989(3) | 1.980(2) | 1.982(2) |
| | O2 x2 | 1.991(2) | 1.967(3) | 1.999(5) | 1.991(3) | 1.988(3) | 1.988(2) | 1.984(3) | 1.974(3) | 1.975(3) | 1.973(2) | 1.971(4) |
| Fe-O-Fe | | 157.2(3) | 156.8(4) | | 158.5(3) | 165.3(4) | 160.1(3) | 161.1(4) | 161.4(3) | 163.1(4) | 163.0(4) | 165.4(4) |
| | | 158.6(4) | 160.2(3) | | 160.2(3) | 164.5(3) | 162.4(4) | 163.6(3) | 165.2(4) | 166.3(3) | 167.2(4) | 168.2(4) |

**Table 6** Polyhedral analysis of LFPTO samples at 295K (x-concentration of PbTiO$_3$, δ – cation shift from centroid, ξ- average bond distance and bond-length variance, V- polyhedral volume, Δ - polyhedral volume distortion).

| Cation | x | δ(Å) | ξ (Å) | V(Å$^3$) | Δ | Valence |
|---|---|---|---|---|---|---|
| La/Pb (c.n.=12) | 0 | 0.194 | 2.800+/-0.324 | 51.40 | 0.072 | 2.96 |
| | 0.1 | 0.125 | 2.794+/-0.284 | 51.37 | 0.070 | 2.94/1.95 |
| | 0.2 | 0.077 | 2.791+/-0.245 | 51.27 | 0.069 | 2.95/1.94 |
| | 0.3 | 0.038 | 2.789+/-0.224 | 51.18 | 0.069 | 2.92/1.93 |
| | 0.4 | 0.013 | 2.787+/-0.194 | 51.17 | 0.068 | 2.93/1.92 |
| | 0.5 | 0.004 | 2.789+/-0.180 | 50.95 | 0.077 | 2.91/1.89 |
| | 0.6 | 0.0008 | 2.788+/-0.154 | 50.94 | 0.077 | 2.89/1.87 |
| | 0.7 | 0.0005 | 2.782+/-0.063 | 50.87 | 0.069 | 2.92/1.93 |
| | 0.8 | 0.262 | 2.800+/-0.168 | 51.31 | 0.071 | 2.98/2.03 |
| | 0.9 | 0.359 | 2.817+/-0.229 | 51.86 | 0.072 | 2.95/2.09 |
| | 1.0 | 0.482 | 2.841+/-0.305 | 52.47 | 0.073 | 2.02 |
| Fe/Ti (c.n.=6) | 0 | 0 | 2.008+/-0.007 | 10.79 | 0 | 2.96 |
| | 0.1 | 0 | 1.999+/-0.027 | 10.65 | 0 | 2.91/3.93 |
| | 0.2 | 0 | 1.992+/-0.047 | 10.53 | 0.001 | 2.86/3.89 |
| | 0.3 | 0 | 1.989+/-0.034 | 10.48 | 0.001 | 2.92/3.92 |
| | 0.4 | 0 | 1.984+/-0.019 | 10.40 | 0.001 | 2.88/3.89 |
| | 0.5 | 0 | 1.983+/-0.023 | 10.33 | 0.007 | 2.90/3.91 |
| | 0.6 | 0 | 1.980+/-0.014 | 10.28 | 0.008 | 2.96/3.94 |
| | 0.7 | 0 | 1.968+/-0.006 | 10.16 | 0.001 | 2.91/3.90 |
| | 0.8 | 0.233 | 1.985+/-0.128 | 10.26 | 0.001 | 2.94/3.96 |
| | 0.9 | 0.269 | 1.995+/-0.156 | 10.37 | 0.001 | 2.96/3.94 |
| | 1.0 | 0.324 | 2.009+/-0.197 | 10.49 | 0.002 | 3.93 |

**Table 7** Basis vectors for the Fe cation for **k=(0,0,0)** in the space group *Pnma*.
The coefficients u,v,w are fixed by the symmetry for one set but are independent for different sites.

|  | $\psi_1$ | $\psi_2$ | $\psi_3$ | $\psi_4$ |
|---|---|---|---|---|
|  | (x,y,z) | (-x+1/2,-y,z-1/2) | (-x,y+1/2,-z) | (x+1/2,-y+1/2,-z+1/2) |
| $\Gamma_1$ | (u, v, w) | (-u, -v, w) | (-u, v, -w) | (u,-v,-w) |
| $\Gamma_3$ | (u, v, w) | (-u, -v, w) | (u, -v, w) | (-u, v w) |
| $\Gamma_5$ | (u, v, w) | (u, v, -w) | (-u, v, -w) | (-u, v, w) |
| $\Gamma_7$ | (u, v, w) | (u, v, -w) | (-u, v, w) | (u, -v, -w) |

**Table 8** Concentration dependence of orthorhombic distortion (s) and tilting angles for orthorhombic perovskites in LFPTO solid solution system:
$s=(a-c)/(a+c)$, $\Theta$–tilting angle around $[101]_p$, $\varphi$–tilting angle around $[010]_p$, $\phi$–tilting angle around $[111]_p$.

| x | s | $\Theta°$ | $\varphi°$ | $\phi°$ |
|---|---|---|---|---|
| 0 | $7\times10^{-4}$ | 3.04 | 1.02 | 3.56 |
| 0.1 | $1\times10^{-3}$ | 3.69 | 3.35 | 3.96 |
| 0.2 | $1.8\times10^{-3}$ | 4.92 | 4.78 | 6.87 |
| 0.3 | $2.5\times10^{-3}$ | 5.83 | 6.09 | 8.43 |
| 0.4 | $2.7\times10^{-3}$ | 5.93 | 6.49 | 8.80 |
| 0.5 | $2.1\times10^{-3}$ | 5.20 | 2.83 | 5.92 |
| 0.6 | $1.2\times10^{-3}$ | 3.99 | 1.98 | 4.11 |
| 0.7 | $2\times10^{-4}$ | 1.65 | 1.96 | 1.07 |

**List of captions for illustrations**

Figure 1 . Temperature dependence of the field cooled (FC) magnetization (H = 1000 or 2000 Oe) for the pure LaFeO$_3$ compound (x=0) and LFPTO solid solutions with x=0.1, 0.2, 0.3, and 0.4. The zero-field cooled (ZFC) magnetization of the pure LaFeO$_3$ compound data is also included for reference. Arrows mark features in the magnetization curves related to the antiferromagnetic ordering of the compounds.

Figure 2. Temperature dependence of zero-field cooled (ZFC), field-cooled (FC), and thermoremanent (TRM) magnetization (H = 20 Oe) for the LFPTO solid solutions with x=0.4, 0.5, and 0.6. The inset shows the temperature dependence of the heat capacity C for the compound with x=0.4.

Figure 3. Magnetic field dependence of the magnetization at low temperatures (T=10K) for the LFPTO solid solutions with x=0.4, 0.5, and 0.6. Results for x=0.7 (main frame) and x=0 (inset, which also includes data recorded at 300K) are included for comparison.

Figure 4. Temperature dependence of dielectric constant $\varepsilon$ for several LPFTO samples with x=1 (a), 0.9 (b) 0.8 (c), 0.7 (d), 0.6 (e), 0.5(f), 0.4 (g), and 0.3 (h) at different frequencies: 100 Hz (1), 1 kHz (2), 10 kHz (3), 100 kHz (4), 1 MHz (5).

Figure 5. Temperature dependence of dielectric loss tgδ for LPFTO samples (x=0.1 (a) and x=0.2 (b)) at different frequencies: 100 Hz (1), 1 kHz (2), 10 kHz (3), 100 kHz (4), 1 MHz (5).

Figure 6.a. Mössbauer spectrum of LFPTO solid solution with x=0.8 recorded at 295 K. The dots are the datapoints, the solid line the fitted Lorentzian doublet.
b. Mössbauer spectrum of LFPTO solid solution with x=0.3 recorded at 295 K, The dots are the datapoints, the solid line is the sum of the fitted Lorentzian functons (3 broad sextets and one doublet).

Figure 7 a. Fitted Mössbauer spectrum for LFPTO solid solution with x=0.5 recorded at room temperature after long time annealing.
b. Fitted Mössbauer spectrum for LFPTO solid solution with x=0.5 recorded at 78 K.

Figure 8 a. Averaged hyperfine field (T) for LFPTO solid solutions. For x = 0.4 and 0.5 two distinct patterns are achieved, one doublet (H = 0 T) and one narrow sextet with H ≈51 T.
b. The intensity of the PM doublet (in %), measured at room temperature versus composition x.

Figure 9. a. Evolution of the XRPD patterns of LFPTO with concentration x.
b. XRPD pattern for LFPTO with x=0.4 at room temperature.

Figure 10. Evolution of lattice parameters of LFPTO with concentration x. For easy comparison, the lattice parameters of the orthorhombic phase are plotted in units of a/$\sqrt{2}$, b/2 and c/$\sqrt{2}$, respectively.

Figure 11. The observed, calculated, and difference plots for the fit to NPD patterns of LFPTO with x=0.2 after Rietveld refinement of the nuclear and magnetic structure at different temperatures: 295 K (a) and 10 K (b).

Figure 12. Evolution of the NPD patterns of LFPTO with x=0.1 with temperature.

Figure 13. A sketch of the magnetic structure of antiferromagnetic LFPTO at T = 10 K. Red arrows represent the magnetic moments of the magnetic cations (blue spheres).

Figure 14. Projection in the ac-plane of the polyhedral representation of the structure of LFPTO for x=0 and x=0.4. Note how the *A*-site cations are shifted for x=0.

Figure 15. Schematic electronic phase diagram summarizing the evolution of structural, dielectric, and magnetic properties of the (1-x) LaFeO$_3$ – (x)PbTiO$_3$ system as a function of the fractional composition. $T_N$ (red circles) is the antiferromagnetic transition temperature, while $T_C$ (blue squares) denotes the ferroelectric one. Open and filled black diamonds reflect the frequency-dependent maximum in the dielectric constants associated with relaxor behavior (f=100 Hz (filled symbols) or 100 kHz (open symbols)). Error bars on critical temperatures (~5K) are omitted for clarity. The dielectric properties of the solutions with x=0.10 and 0.20 could not be investigated. "O" and "T" refer to low-temperatures orthorhombic and tetragonal structures, respectively. The orthorhombic strain s in the "O" phase is plotted in green triangles (right axis). The shaded area reflects the possible position of morphotropic phase boundary (see main text for details).